\documentstyle[preprint,aps,epsf]{revtex}
\begin{document}
\draft
\preprint{FNT/T 97/08, MKPH-T-97-5}
\title{Treatment of the $\Delta$ current in electromagnetic 
two-nucleon knockout reactions}
\author{P.\ Wilhelm and H.\ Arenh\"ovel}
\address
{Institut f\"ur Kernphysik, Johannes Gutenberg-Universit\"at, D-55099 Mainz, 
Germany}
\author{C.\ Giusti and F.\ D.\ Pacati}
\address
{Dipartimento di Fisica Nucleare e Teorica, Universit\`a di Pavia, and\\
Istituto Nazionale di Fisica Nucleare, Sezione di Pavia, Italy}
\date{\today}
\maketitle
\begin{abstract}
  The contribution of the $\Delta$(1232) isobar to the electromagnetic
  current of the two-nucleon system and its role in ($\gamma$,NN)
  processes is investigated.  The difference between the genuine
  $\Delta$-excitation current and that part of the current connected
  to the deexcitation of a preformed $\Delta$ in the target nucleus is
  stressed. The latter cannot lead to a resonant behaviour of matrix
  elements for energies in the $\Delta$ region. The reaction
  $^{16}\mbox{O}(\gamma,pp)^{14}\mbox{C}$, where the $\Delta$
  contribution is dominant at intermediate energies, is considered.
  The large variations found in the cross sections for different
  treatments stress the need for a proper treatment of the $\Delta$
  current for a clear understanding of the reaction mechanism of
  two-nucleon emission processes.
\end{abstract}

\section{Introduction}

The electromagnetically induced two-nucleon knockout serves as a tool
to study short-range correlations between two nucleons in a nucleus.
Thereby, one assumes that the photon interacts with the correlated
pair through a one-nucleon current.  However, nucleon pairs can also
be ejected by two-nucleon currents which effectively take into account
the influence of subnuclear degrees of freedom like mesons and
isobars.  Therefore, in order to estimate this competing effect
quantitatively, a reliable treatment of meson exchange as well as
isobar currents is necessary before one can draw definite conclusions
on the role of short-range correlations.  Experimental data of
two-nucleon photoemission in the $\Delta(1232)$ region were taken in
different laboratories \cite{Kana,Are,Audi,Cro}, while only one
exploratory experiment on the $(e,e^\prime pp)$ reaction \cite{Zonde} was
completed. However, many new results will become available in the near
future, exploiting continuous-wave accelerators and tagged-photon
facilities.

\begin{figure}
\centerline{\epsfxsize=12cm\epsffile{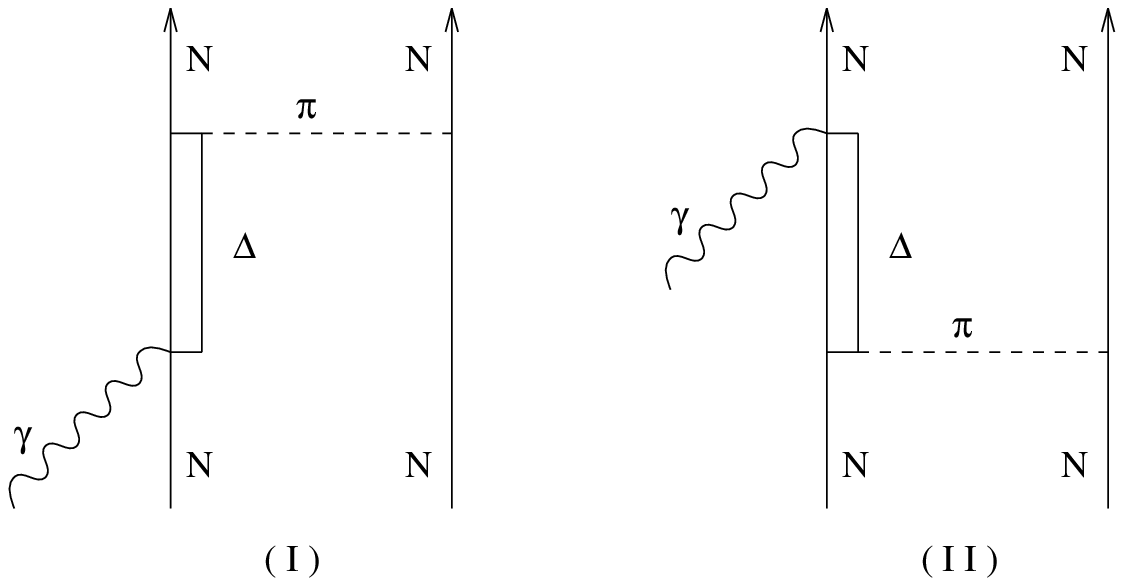}}
\caption{
  The $\Delta$ contribution to the current of the two-nucleon system:
  $\Delta$-excitation current (I) and $\Delta$-deexcitation current
  (II).
\label{fig:graph}
}
\end{figure}

In this note we would like to point out a pitfall which one may
encounter when considering contributions from intermediate $\Delta$
isobars to electromagnetic processes via the effective operator
approach.  To this end we remind the reader at the two first-order
contributions of the $\Delta$ to the effective electromagnetic
two-body current of a two-nucleon subsystem shown diagramatically in
Fig.\ \ref{fig:graph}.  The first one (I) describes the $\Delta$
excitation with subsequent deexcitation by pion exchange while the
second (II) describes the time interchange of the two steps, i.e.,
first excitation of a virtual $\Delta$ by pion exchange in a NN
collision and subsequent deexcitation by photon absorption.  In other
words parts I and II correspond to N$\Delta$ admixtures in the final
and the initial NN subsystem, respectively. They would be treated as
explicit components of the wave function in the approach of nuclear
isobar configurations (IC) \cite{green,hamainz}.  Consequently, it is
important to note that in diagram II the $\Delta$ is always far
off-shell being part of the initial state irrespective of the energy
transferred to the system by the real (or virtual) photon, whereas in
diagram I the $\Delta$ can become on-shell for a sufficiently high
energy transfer. This then gives rise to a pronounced resonant
behaviour of the matrix elements of diagram I when varying the energy
transfer in the region between, say, 200$\,$ and 400$\,$MeV.  These
features appear automatically in the IC approach, however, in the
effective operator approach only if one keeps the full $\Delta$
propagator in the intermediate state, making the effective two-body
operator nonlocal.

For ($\gamma$,NN) calculations on complex nuclei, it is necessary to
avoid this nonlocality in order to keep the numerical effort within
reasonable limits.  Therefore, often the $\Delta$ propagator is taken
in the simplest static approximation keeping only the baryon mass
difference $M_\Delta-M_N$.  In this case the two contributions of
Fig.\ \ref{fig:graph} can be combined into one effective local
two-nucleon operator. This approximation appears reasonable at low
energies but it certainly fails when the transferred energy allows the
excitation of an on-shell $\Delta$.  In the latter case one might be
tempted to replace $(M_\Delta-M_N)^{-1}$ by a resonant
energy-dependent $\Delta$ propagator.  Indeed this procedure has been
followed in the past.  However, it leads to a wrong effective operator, 
as is outlined in detail in the next section, and results in a strong
overestimation of the contribution of diagram II.  This is shown in
section \ref{results}, where different treatments of the $\Delta$
current are compared by means of a calculation for the
$^{16}\mbox{O}(\gamma,pp)^{14}\mbox{C}$ process within the framework
of Refs.\ \cite{boffi,giusti}.

\section{Formalism} \label{formalism}

The effective current operator shown in Fig.\ \ref{fig:graph} is given
by
\begin{equation}
  \vec\jmath_\Delta = \vec\jmath_\Delta^{\,(I)} +
  \vec\jmath_\Delta^{\,(II)} +(1\leftrightarrow 2).
\label{eq:j12}
\end{equation}
In the following we restrict ourselves to the dominant magnetic dipole
N$\leftrightarrow\!\Delta$ transition. For simplicity the hadronic
N$\Delta\!\leftrightarrow$NN transition is described by static
$\pi$-exchange. The inclusion of the $\rho$-exchange is
straightforward but not essential for our purpose.  Under these
assumptions, the excitation current reads
\begin{equation}
  \vec\jmath_\Delta^{\,(I)}(\vec q) = \gamma\,
  \vec\tau^{(1)}_{N\Delta}\cdot\vec\tau^{(2)}_{NN}\,
  \frac{\vec\sigma^{(1)}_{N\Delta}\cdot\vec
    k\,\vec\sigma^{(2)}_{NN}\cdot\vec k} {\vec k^2+m_\pi^2}
  \,\,G_\Delta(\sqrt{s_{I}})\,\, \vec\tau^{(1)}_{\Delta N,3}\,\,
  i\vec\sigma^{(1)}_{\Delta N}\times\vec q,
\label{eq:j1}
\end{equation}
and the deexcitation part
\begin{equation}
  \vec\jmath_\Delta^{\,(II)}(\vec q) = \gamma\,
  \vec\tau^{(1)}_{N\Delta,3}\,\, i\vec\sigma^{(1)}_{N\Delta}\times\vec
  q \,\,G_\Delta(\sqrt{s_{II}})\,\, \vec\tau^{(1)}_{\Delta
    N}\cdot\vec\tau^{(2)}_{NN}\, \frac{\vec\sigma^{(1)}_{\Delta
      N}\cdot\vec k\,\vec\sigma^{(2)}_{NN}\cdot\vec k} {\vec
    k^2+m_\pi^2},
\label{eq:j2}
\end{equation}
where $\vec q$ is the photon momentum, and $\vec k$ is the momentum of
the exchanged pion.  The factor $\gamma$ collects various coupling
constants, $\gamma=f_{\gamma N\Delta}f_{\pi NN}f_{\pi
  N\Delta}/m_\pi^3$.  The propagator of the resonance is denoted by
$G_\Delta$.  It depends on the invariant energy $\sqrt s$ of the
$\Delta$.  
Neglecting the kinetic energy of the relative motion of the intermediate 
N$\Delta$ state, one obtains a
local approximation to $G_\Delta$. It reads
\begin{equation}
  G_\Delta(\sqrt s)=\frac{1} {{M_\Delta}-\sqrt
    s-\frac{i}{2}\Gamma_\Delta(\sqrt s)},
\label{eq:prop}
\end{equation}
where $\Gamma_\Delta$ is the energy-dependent decay width of the
$\Delta$ and $M_\Delta=1232\,$MeV its mass.

Clearly, $\sqrt s$ can be very different in diagram I and II. In
diagram II, it does not depend on the photon energy $E_\gamma$, and
it is reasonable to approximate it by the nucleon mass
\begin{equation}
  \sqrt{s_{II}} = M_N.
\label{eq:s2}
\end{equation}
On the other hand, $\sqrt{s_{I}}$ depends on $E_\gamma$.  It grows
with $E_\gamma$ and for $\sqrt{s_I} \approx M_\Delta$ the $\Delta$ is
essentially on-shell.  Following the recent suggestion made in
\cite{comment} for the choice of $\sqrt{s_I}$, the calculations
presented below use
\begin{equation}
  \sqrt{s_{I}}=\sqrt{s_{NN}}-M_N,
\label{eq:s1}
\end{equation}
where $\sqrt{s_{NN}}$ is the experimentally measured invariant energy
of the two fast outgoing nucleons in an $A(\gamma,NN)A-2$ reaction.
The energy dependence of the two parts of the $\Delta$ current can
also be considered from a different point of view.  For forward
propagating exchanged pions, both parts are related to the process
of electromagnetic pion production on one nucleon followed by its
reabsorption on the second nucleon.  Part I is connected to the
$s$-channel contribution of the $\Delta$ which leads to the well-known
resonant $M_{1+}^{3/2}$ pion production multipole, whereas part II is
connected to the $u$-channel contribution which has only a smooth energy
dependence.

Only for low energy transfers, say below 100$\,$MeV, it may be
justified to approximate also $\sqrt{s_{I}}\approx M_N$. Then the
$\Delta$ propagators in the excitation and deexcitation parts are equal
and the spin and isospin structure of their sum simplifies due to the
cancellation of terms.  To see this, one first has to rewrite Eqs.\ 
(\ref{eq:j1}) and (\ref{eq:j2}) using the following identity for the
N$\leftrightarrow\!\Delta$ transition spin (isospin) operators 
\begin{equation}
  \label{eq:id}
  \vec\sigma_{N\Delta}\cdot\vec a\,\, 
  \vec\sigma_{\Delta N}\cdot\vec b = \frac{2}{3}\vec a\cdot\vec b
  -\frac{i}{3}\vec\sigma_{NN}\cdot\vec a\times\vec b,
\end{equation}
where $\vec a$ and $\vec b$ are two arbitrary vectors.  One finds
\begin{eqnarray}
  &\vec\jmath_\Delta^{\,(I)}(\vec q) =& \frac{1}{9}\,\gamma\, \left[
    2\vec\tau^{(2)}_{NN,3} -
    i\left(\vec\tau^{(1)}_{NN}\times\vec\tau^{(2)}_{NN}\right)_3
  \right] \nonumber\\ && \left( 2i\vec k - \vec k \times 
   \vec\sigma^{(1)}_{NN} \right) \times \vec q \,\, G_\Delta(\sqrt{s_{I}})\,\,
  \frac{\vec\sigma^{(2)}_{NN}\cdot\vec k}{\vec k^2+m_\pi^2}
\label{eq:j1s}
\end{eqnarray}
and 
\begin{eqnarray}
  &\vec\jmath_\Delta^{\,(II)}(\vec q) =& \frac{1}{9}\,\gamma\, \left[
    2\vec\tau^{(2)}_{NN,3} +
    i\left(\vec\tau^{(1)}_{NN}\times\vec\tau^{(2)}_{NN}\right)_3
  \right] \nonumber\\ && \left( 2i\vec k + \vec k \times 
   \vec\sigma^{(1)}_{NN} \right) \times \vec q \,\, G_\Delta(\sqrt{s_{II}})\,\,
  \frac{\vec\sigma^{(2)}_{NN}\cdot\vec k}{\vec k^2+m_\pi^2}.
\label{eq:j2s}
\end{eqnarray}
Then, in the low-energy (le) approximation, i.e.\ using
$G_\Delta(\sqrt{s_{I}})=G_\Delta(\sqrt{s_{II}})=(M_\Delta-M_N)^{-1}$,
one obtains for the total current
\begin{eqnarray}
  &\vec\jmath^{\,({\mathrm{le}})}_\Delta(\vec q) =& \frac{2}{9}\,\gamma \,i\, 
  \left[ 4\vec\tau^{(2)}_{NN,3}\,\vec k + \left(\vec\tau^{(1)}_{NN}\times
    \vec\tau^{(2)}_{NN}\right)_3 \,
    \vec k\times\vec\sigma^{(1)}_{NN} \right]\times \vec q
  \nonumber\\ && \frac{1}{M_\Delta-M_N}\,
  \frac{\vec\sigma^{(2)}_{NN}\cdot\vec k}{\vec k^2+m_\pi^2}.
\label{eq:le}
\end{eqnarray}
This form is usually quoted in the literature (see e.g.\ 
\cite{riska,orden}).  It serves as starting point for model
calculations in heavier nuclei \cite{orden,alberico,pavia} as well as
in few-nucleon reactions \cite{hockert,tamura,mosconi}.

The simple replacement of the low energy propagator
$(M_\Delta-M_N)^{-1}$ in Eq.\ (\ref{eq:le}) by the resonant $G_\Delta$
of Eq.\ (\ref{eq:prop}) in order to obtain an operator which is more
appropriate for studying photon absorption at higher energies is,
however, in no way justified.  This procedure implies a strong
overestimation of the deexcitation part at higher energies since it
introduces a resonant behaviour also there.  Nevertheless, it has been
used in the past (see e.g.\ \cite{dekker,ryckebusch}).  To make this
point clearer, it is useful to consider the isospin matrix structure
of the excitation and deexcitation current.  Table \ref{tab:iso}
summarizes the relevant isospin matrix elements.  Note that
transitions between $np$ pairs with isospin $T=1$ are always forbidden
for an isovector current.  According to Table \ref{tab:iso}, the above
replacement leads to wrong conclusions, in particular when the
absorption on $pp$ pairs is studied, or in cases where the absorption
on $np$ pairs with isospin $T=1$ is expected to be relevant.  In
($\gamma,np$) reactions, only those contributions which proceed via
absorption on $np$ pairs with $T=0$ (like the quasi deuteron mechanism
for example) remain unaffected with respect to the correct
prescription, as the contribution of the operator (\ref{eq:j2s})
vanishes.

\section{Results} \label{results}

In this section we investigate the dependence of $(\gamma,pp)$ cross
sections on the different treatments of the $\Delta$ current
within the theoretical model of Ref.\ \cite{giusti}. Although different
channels can be considered in the model, we here have chosen the
$(\gamma,pp)$ channel since there the $\Delta$ current is dominant at
intermediate energies.

Exclusive cross sections of the $^{16}\mbox{O}(\gamma,pp)^{14}\mbox{C}$
knockout reaction have been calculated for transitions to the ground state and
low-lying discrete excited states of $^{14}\mbox{C}$. In the model the
final state $|J^{\pi}\rangle$ 
of the residual nucleus is obtained from the removal of a
nucleon pair coupled to $J$. The correlated wave function of
the pair is calculated with the single-particle states of Ref.\ \cite{elton}
and the Jastrow-type correlation function of Ref.\ \cite{gear}. The
final-state interaction is taken into account by means of the optical
potential of Ref.\ \cite{nada}, describing the interaction of each one of the
two protons with the residual nucleus. More details of the model and of the 
theoretical ingredients of the calculation are given in Ref.\ \cite{giusti}.

\begin{figure}
\centerline{\epsfxsize=12cm\epsffile{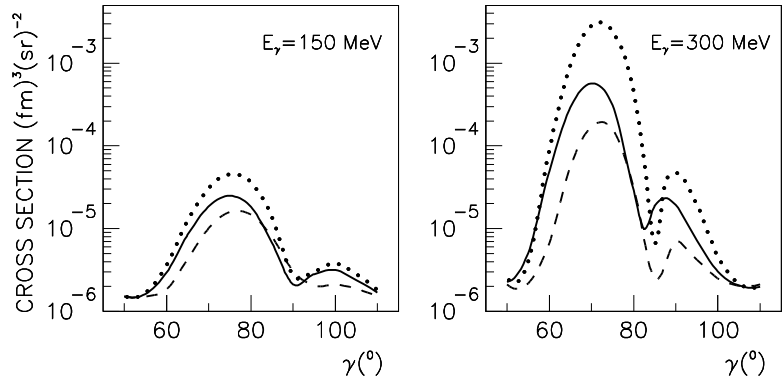}}
\caption{
  The differential cross section of the
  $^{16}\mbox{O}(\gamma,pp)^{14}\mbox{C}(g.s.)$ reaction in coplanar
  and symmetrical kinematics as a function of $\gamma$ using different
  treatments of the $\Delta$ current: $\vec\jmath_\Delta$(RN) solid,
  $\vec\jmath_\Delta$(RR) dotted, and $\vec\jmath_\Delta$(NN) dashed
  curves.
\label{fig:gamma}
}
\end{figure}

The differential cross sections  of the 
$^{16}\mbox{O}(\gamma,pp)^{14}\mbox{C}(g.s.)$ reaction at $E_\gamma=150\,$MeV 
and $300\,$MeV are shown in Fig.\ 
\ref{fig:gamma}  in a coplanar and symmetrical
kinematics as a function of the angle $\gamma$ between the photon and
one of the symmetrically outgoing protons. 
Three different treatments of the $\Delta$ current are
compared.  The solid curves refer to the correct form, i.e., use of
the resonant $\Delta$ propagator in the excitation current only (part
I of Fig.\ \ref{fig:graph}).  Adopting the resonant propagator in both
excitation and deexcitation currents (part I and II) leads to the
dotted curves.  The dashed curves refer to the low energy
approximation of Eq.\ (\ref{eq:le}). In the following, these three
currents are referred to as $\vec\jmath_\Delta$(RN),
$\vec\jmath_\Delta$(RR), and $\vec\jmath_\Delta$(NN), respectively,
indicating the use of either the resonant (R) or energy-independent
nonresonant (N) propagator in part I and II.  Already at 150$\,$MeV,
$\vec\jmath_\Delta$(RN) and $\vec\jmath_\Delta$(RR) lead to peak cross
sections which differ by nearly a factor two. The difference grows
with the photon energy and at 300$\,$MeV the peak cross sections
differ by more than a factor five.  As one would expect,
$\vec\jmath_\Delta$(NN) underestimates and $\vec\jmath_\Delta$(RR) 
overestimates the cross section.  
One notes also a different shape of the angular
distributions for $\vec\jmath_\Delta$(RN) on the one hand and
$\vec\jmath_\Delta$(RR) or $\vec\jmath_\Delta$(NN) on the other hand.
At 300$\,$MeV the minimum is less pronounced for the solid curve.  This is a
consequence of the fact that $\vec\jmath_\Delta$(RN) has a different
operator structure with respect to its spin and angular momentum
dependence, since its parts I and II do not enter with equal weight as
in $\vec\jmath_\Delta$(RR) and $\vec\jmath_\Delta$(NN).  For the same
reason one may expect qualitatively different predictions for
polarization observables.
 
\begin{figure}
\centerline{\epsfxsize=12cm\epsffile{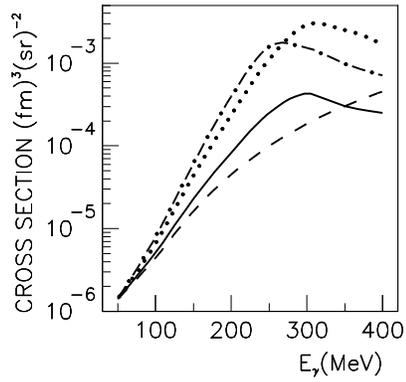}}
\caption{
  The differential cross section of the
  $^{16}\mbox{O}(\gamma,pp)^{14}\mbox{C}(g.s.)$ reaction in coplanar
  and symmetrical kinematics at zero recoil momentum as a function of the
  photon energy.  Line convention as in Fig.\ \protect\ref{fig:gamma}.
  The dash-dotted curve has been calculated with
  $\vec\jmath_\Delta$(RR), but using the energy assignment
  $s_I^{1/2}=E_\gamma+M_N$ as in Ref.\ \protect\cite{ryckebusch}.
\label{fig:energy}
}
\end{figure}

In Fig.\ \ref{fig:energy} the differential cross section in coplanar
and symmetrical kinematics at zero recoil momentum of the residual
nucleus is plotted as a function of the photon energy.  This
corresponds to the region in Fig.\ \ref{fig:gamma} where the cross
section is maximal.  Fig.\ \ref{fig:energy} clearly shows the strong
overestimation of the cross section when $\vec\jmath_\Delta$(RR) is
used. This certainly would affect any analysis of experimental
$(\gamma,pp)$ data in the $\Delta$ region in view of the role of
short-range correlations.  The low-energy approximation
$\vec\jmath_\Delta$(NN) can of course not predict a resonance peak.
The dash-dotted curve has been calculated with $\vec\jmath_\Delta$(RR) as the
dotted one, but the choice in Eq.\ (\protect\ref{eq:s1}) has been replaced by
the energy assignment $\sqrt{s_I}=E_\gamma+M_N$ used, e.g., in
Ref.\ \cite{ryckebusch}. It shifts the resonance position towards lower
energies and thus leads to a further overestimation of the cross
section for energies below 260$\,$MeV.  The effect of this choice of
$\sqrt{s_I}$ on the angular distribution has already been discussed in
Ref. \cite{giusti} for the $(e,e'pp)$ reaction.

Calculations for the transition to the excited $1^+$ and to the first
excited $2^+$ state of $^{14}\mbox{C}$ have also been performed.
Qualitatively, the results agree with the former ones.  However, the
size of the overestimation varies.  This again can be traced back to
the different operator structure of $\vec\jmath_\Delta$(RN) and
$\vec\jmath_\Delta$(RR). It becomes visible when the quantum numbers
of the active nucleon pair change and cannot simply be simulated even
by an energy-dependent rescaling factor.

Similar results are obtained in the $(e,e'pp)$ reaction, where the 
presence  of the longitudinal contribution, due to correlations, reduces the 
effects of the different treatments of the $\Delta$ current.

\section{Summary}

In this paper we have discussed the treatment of the $\Delta$ isobar
current in electromagnetic two-nucleon knockout reactions based on an
effective two-nucleon operator.  It has been emphasized that this
current consists of a $\Delta$ excitation part and a $\Delta$
deexcitation part corresponding to a $\Delta$ admixture in the initial
state.  These two parts are related to the $s$- and $u$-channel
contributions of the $\Delta$ to electromagnetic pion production on a
nucleon.  
Only the excitation part of the current has a resonant
energy dependence which can result in a characteristic energy
dependence of observables.
  
This different energy dependence of excitation and deexcitation parts
has to be taken care of when going to higher energy transfers.  In
particular, it does not allow to replace the propagator
$(M_\Delta-M_N)^{-1}$ of the simplest static approximation to the
current (which is justified in the low-energy region) by a resonant
propagator.  The error is two-fold: first, the deexcitation part would
get a wrong energy dependence and, secondly, one would loose parts of
the excitation current which have been cancelled by terms in the
deexcitation current.  In view of the fact that this prescription has
been used in the literature, we have performed an explicit calculation
for the $^{16}\mbox{O}(\gamma,pp)^{14}\mbox{C}$ reaction.  It shows
that the simple replacement may lead to an overestimation of the cross
section (in the preferred kinematic region around zero recoil momentum
of the residual nucleus) by more than a factor five.  Such a variation
of the $\Delta$ contribution is definitely too large when one wants to
use two-proton knockout as a tool to study short range correlations in
the nucleus.

\section*{Acknowledgements}

P.W.\ thanks the Dipartimento di Fisica Nucleare e Teorica of the
University of Pavia for the hospitality during part of this work. H.A.\ 
and P.W.\ were supported by the Deutsche Forschungsgemeinschaft (SFB
201).


\begin{table}
\caption{Isospin matrix elements 
  (terms in square brackets in Eqs.\ (\protect\ref{eq:j1s}) and
  (\protect\ref{eq:j2s})) of the excitation (I) and deexcitation part
  (II) of the $\Delta$ current for various transitions.}
\label{tab:iso}
\begin{tabular}{lcc}
Transition & $\vec\jmath_\Delta^{\,(I)}$ & $\vec\jmath_\Delta^{\,(II)}$ \\
\tableline
$np(T=0)\rightarrow np(T=1)$ & $-4$ & 0 \\
$np(T=1)\rightarrow np(T=0)$ & 0 & $-4$ \\
$pp/nn\rightarrow pp/nn$     & $\pm 2$ & $\pm 2$ \\
\end{tabular}
\end{table}
\end{document}